\documentclass[aps,superscriptaddress,preprintnumbers,showpacs,twocolumn]{revtex4}
\usepackage{amssymb}
\usepackage{dcolumn}
\usepackage{bm}
\usepackage{graphicx}
\usepackage{booktabs}
\usepackage{multirow}

\begin{document}

\newcommand*{\PKU}{School of Physics and State Key Laboratory of Nuclear Physics and
Technology, Peking University, Beijing 100871,
China}\affiliation{\PKU}
\newcommand*{\CHEP}{Center for High Energy Physics, Peking University, Beijing 100871, China}\affiliation{\CHEP}

\title{Correlating lepton mixing angles and mixing matrix with Wolfenstein parameters}

\author{Xinyi Zhang}\affiliation{\PKU}
\author{Bo-Qiang Ma}\email{mabq@pku.edu.cn}\affiliation{\PKU}\affiliation{\CHEP}

\begin{abstract}
Inspired by a new relation $\theta_{13}^{\rm
PMNS}={\theta_C}/{\sqrt{2}}$ observed from the relatively large
$\theta_{13}^{\rm PMNS}$, we find that the combination of this
relation with the quark-lepton complementarity and the
self-complementarity results in correlations of the lepton
mixing angles with the quark mixing angles. We find that the three
mixing angles in the PMNS matrix are all related to the Wolfenstein
parameter $\lambda$ in the quark mixing, so they are also
correlated. Consequently, the PMNS matrix can be parameterized by
$\lambda$, A, and a Dirac CP-violating phase $\delta$. Such
parametrizations for the PMNS matrix have the same explicitly
hierarchical structure as the Wolfenstein parametrization for the
CKM matrix in the quark mixing, and the bimaximal mixing
pattern is deduced at the leading order. We also discuss
implications of these phenomenological relations in
parametrizations.
\end{abstract}

\maketitle

\section{INTRODUCTION}

Seeking a symmetry or unification of quarks and leptons is one of
the goals of particle physics, and many efforts are devoted to this
area. The bottom-up approach, i.e., finding
some phenomenological relations as well as their explanations, gives
some clues on this issue. The purpose of this paper is to
investigate a number of phenomenological relations between the quark
and lepton mixing parameters and their implications in
phenomenological analysis.

Generally, the mixing of quarks is described by the
Cabibbo-Kobayashi-Maskawa~(CKM) matrix~\cite{CKM} and the mixing of
leptons is described by the Pontecorvo-Maki-Nakagawa-Sakata (PMNS)
matrix~\cite{PMNS}. Both of them are unitary matrices that can be
parametrized by different methods, and phenomenological relations
are expressions of parameters in the context of certain
parametrization. Among many parametrizations, the Chau-Keung (CK)
parametrization~\cite{CK}, i.e., the standard parametrization
adopted by the Particle Data Group~\cite{PDG1996,pdg2010} and the
Wolfenstein parametrization~\cite{Wolfenstein:1983yz} are widely
used for the CKM matrix of quark mixing. The former can be
applied to the analysis of the relevant experiments with three
mixing angles and a CP-violating phase, while the latter is a
good indicator of the hierarchical structure for quark mixing
between different generations.

The mixing angles in the PMNS matrix are generally larger than their
correspondents in the CKM matrix, and this renders the hierarchical
structure in the lepton mixing a subtle subject. In the original
Wolfenstein parametrization, the CKM matrix can be well-described as
an expansion around the unit matrix with an expanding parameter
$\lambda$. Following the same spirit and considering the anarchy of
the lepton mixing, the Wolfenstein-like parametrization for the
PMNS matrix is done around a zeroth order matrix. There are several
mixing patterns that are candidates for the zeroth order, such as the
bimaximal pattern~\cite{bi} and the tribimaximal pattern~\cite{tri}.
Although each one of them has assumed theoretical backgrounds,
nature may have a preference for only one of them. We show in this
paper that the bimaximal pattern can be deduced naturally with the assistance of some phenomenological relations observed from data.

The neutrino oscillation experiments are analyzed in the $3\nu$
framework with parameters $\theta_{12}$, $\theta_{23}$, $\theta_{13}$, and a Dirac CP-violating phase $\delta$ in the mixing
matrix. The Majorana CP-violating phases do not manifest themselves
in the oscillation, and thus they are omitted in the discussion. Since
last year, a number of experiments provided evidence for a
relatively large $\theta_{13}$~\cite{T2K,expt2,expt3}. The Daya-Bay
Collaboration~\cite{An:2012eh} and the RENO
Collaboration~\cite{Ahn:2012nd} have established the relatively
large value of $\theta_{13}$ with significance around $5\sigma$. The
progress in the measured values of the lepton mixing angles makes a
revisit of the phenomenological relations necessary. The following
three kinds of phenomenological relations have attracted our
attention:
\begin{enumerate}
\item The quark-lepton complementarity (QLC)~\cite{smirnov,qlc,raidal},\\
$\theta_{12}^{\rm CKM}+\theta_{12}^{\rm PMNS}\simeq45^\circ$,\\
$\theta_{23}^{\rm CKM}+\theta_{23}^{\rm PMNS}\simeq45^\circ$;
\item The self-complementarity (SC)~\cite{Zheng:2011uz,Zhang:2012xu},\\
$\theta_{12}^{\rm PMNS}+\theta_{13}^{\rm PMNS}\simeq\theta_{23}^{\rm
PMNS}$;
\item A new relation, $\theta_{13}^{\rm
PMNS}\simeq\frac{\theta_C}{\sqrt{2}}$.
\end{enumerate}
All four equations are admissible within $3\sigma$. Taking three
of the four equations as valid simultaneously, all the mixing angles
in the lepton sector can be related to the same one ($\lambda$) or
two ($\lambda$, $A$) quark mixing parameters. The correlations of
the lepton mixing angles reduce the number of free parameters in the
PMNS matrix. Thus we arrive at a simplified Wolfenstein
parametrization for the PMNS matrix with parameters $\lambda$, $A$
coming from the quark mixing and a phase representing the CP
violation.

Seeking underlying connections between quarks and leptons is one
of the basic pursuits of flavor physics. Phenomenologically and
historically, there are many relations that have emerged as possible
candidates. The three kinds of relations we discuss here are all
in agreement with the current data. Exploring their implications may
inspire investigation in the direction of model building.

The paper is organized as follows. In Sec.~\ref{sec:relation} we
first check the validity of the above relations with the latest
data. Then we discuss each relation with its origination and
implications in detail. In Sec.~\ref{sec:correlation} we perform the
Wolfenstein parametrization in four correlating scenarios indicated
by the relations. Some discussions and conclusions are presented in
Sec.~\ref{sec:discuss}.

\section{\label{sec:relation}PHENOMENOLOGICAL RELATIONS OF QUARK AND LEPTON MIXING ANGLES}

\subsection{Verification with the current data}

Since the relations under discussion are not all new ones indicated by the current data, examining them with the latest data is necessary.

For the mixing angles in the CKM matrix, we use the global fit result of the Wolfenstein parameters~\cite{pdg2010} and the following reparametrization relations:
\begin{eqnarray}
s_{12}^{\rm CKM}&=&\lambda=0.2253\pm0.0007,\nonumber\\
s_{23}^{\rm CKM}&=&A \lambda^2=0.0410^{+0.0011}_{-0.0008},\nonumber\\
s_{13}^{\rm CKM}&=&|A \lambda^3(\rho+i \eta)|=0.0034^{+0.0002}_{-0.0001}.
\end{eqnarray}
The corresponding mixing angles are
\begin{eqnarray}
\theta_{12}^{\rm CKM}&=&(13.02\pm0.041)^\circ,\nonumber\\
\theta_{23}^{\rm CKM}&=&(2.35^{+0.063}_{-0.046})^\circ,\nonumber\\
\theta_{13}^{\rm CKM}&=&(0.19^{+0.011}_{-0.006})^\circ.
\end{eqnarray}

For the mixing angles in the PMNS matrix, we use the latest global fit result~\cite{Fogli:2012ua} in Table~\ref{tab:globalfit}.
\begin{table}
\caption{\label{tab:globalfit} The recent global fit result of
mixing angles~\cite{Fogli:2012ua}.}
\begin{ruledtabular}
 \begin{tabular}{ccc}
   \toprule
   Parameter& NH &IH\\
   \hline
   $\sin^2\theta_{12}/10^{-1}$&$3.07^{+0.18}_{-0.16}$&$3.07^{+0.18}_{-0.16}$\\
   $\sin^2\theta_{23}/10^{-1}$&$3.98^{+0.30}_{-0.26}$&$4.08^{+0.35}_{-0.30}$\\
   $\sin^2\theta_{13}/10^{-2}$&$2.45^{+0.34}_{-0.31}$&$2.46^{+0.34}_{-0.31}$\\
   \bottomrule
   \end{tabular}
\end{ruledtabular}
\end{table}
The corresponding mixing angles are
\begin{eqnarray}
\theta_{12}^{\rm PMNS}&=&(33.65^{+1.12}_{-0.99})^\circ(\rm NH,\rm IH),\nonumber\\
\theta_{23}^{\rm PMNS}&=&(39.11^{+1.76}_{-1.52})^\circ(\rm NH),(39.70^{+2.04}_{-1.75})^\circ(\rm IH),\nonumber\\
\theta_{13}^{\rm PMNS}&=&(9.01^{+0.63}_{-0.57})^\circ(\rm NH),(9.02^{+0.63}_{-0.57})^\circ(\rm IH),
\end{eqnarray}
where NH/IH denotes normal/inverted hierarchy.

Now we are ready to check the relations we are about to use.

We notice that-although the smallest lepton mixing angle $\theta_{13}^{\rm PMNS}$ turns out to be much larger than expected, which means a reallocation of the values of the other two mixing angles-the quark-lepton complementarity still approximately holds. By observing
\begin{eqnarray}
\theta_{12}^{\rm CKM}+\theta_{12}^{\rm PMNS}&=&(46.67^{+1.16}_{-1.03})^\circ(\rm NH,\rm IH),\nonumber\\
\theta_{23}^{\rm CKM}+\theta_{23}^{\rm PMNS}&=&(41.46^{+1.82}_{-1.57})^\circ(\rm NH)\nonumber\\
&=&(42.05^{+2.10}_{-1.80})^\circ(\rm IH),
\end{eqnarray}
we find that $45^\circ$ lies in the $2\sigma$ range.

For the self-complementarity relation,
\begin{eqnarray}
\theta_{12}^{\rm PMNS}+\theta_{13}^{\rm PMNS}&=&(42.66^{+1.75}_{-1.56})^\circ(\rm NH)\nonumber\\
&=&(42.67^{+1.75}_{-1.56})^\circ(\rm IH),\\
\theta_{23}^{\rm PMNS}&=&(39.11^{+1.76}_{-1.52})^\circ(\rm NH)\nonumber\\
&=&(39.70^{+2.04}_{-1.75})^\circ(\rm IH),
\end{eqnarray}
which makes $\theta_{12}^{\rm PMNS}+\theta_{13}^{\rm PMNS}\simeq\theta_{23}^{\rm PMNS}$ valid in the $2\sigma$ range.

Notice that
\begin{eqnarray}
&&\sin\theta_C=\sin\theta_{12}^{\rm CKM}=\lambda=0.2253\pm0.0007,\\
&&\frac{\lambda}{\sqrt{2}}=0.159\pm0.001,\\
&&\sin\theta_{13}^{\rm PMNS}=0.157^{+0.011}_{-0.010},
\end{eqnarray}
which makes $\sin\theta_{13}^{\rm PMNS}\simeq\frac{\lambda}{\sqrt{2}}$ valid in the $1
\sigma$ range.

The relations are all in agreement with the experimental data at
least within the $2\sigma$ error range. Although taking all the
relations as true is restrictive, it is still beneficial to see the
consequences in such scenarios with as many relations as possible.
Before doing that, we firstly discuss these kinds of relations
separately, focusing on their phenomenological indications,
and especially on revealing the Wolfenstein-like parametrization of the
PMNS matrix.

\subsection{The quark-lepton complementarity (QLC)}

The relation $\theta^{\rm PMNS}_{12}+\theta^{\rm
CKM}_{12}\simeq45^\circ$ was first noted by
Smirnov~\cite{smirnov}. Raidal also noticed the second
complementarity relation $\theta^{\rm PMNS}_{23}+\theta^{\rm
CKM}_{23}\simeq45^\circ$ as experimental evidence for grand
unification~\cite{raidal}. The name, i.e., ``the quark-lepton
complementarity", is firstly proposed in Ref~\cite{qlc}.

As free parameters in the standard model, the mixing angles of the
CKM matrix and the PMNS matrix are theoretically independent of each other. The quark-lepton complementarity offers a promising
possibility for linking the mixing matrices of quarks and
leptons~\cite{Li:2005ir}. Its theoretical realization has been discussed, e.g., in dihedral flavor symmetry~\cite{dihedral}. Its phenomenological implications have also been widely discussed~\cite{phenomenology}.

The quark-lepton complementarity is a numerical relation between the
mixing angles of corresponding mixing matrices. The mixing angles
are the parameters in the standard parametrization. Considering the
equivalence of different parametrizations, the relation, reflecting
the correlation between the two mixing matrices in nature, can be
adopted for different parametrizations~\cite{Zheng10,Zhang:2012zh}.

The triminimal parametrization~\cite{triminimal} was first used
in the lepton mixing as an expansion in the correction of mixing
angles. It shares the advantage of having a clear physical meaning
with the standard parametrization and the advantage of possessing
the hierarchical structure with the Wolfenstein
parametrization~\cite{He:2008td,He:2009jm}. The zeroth-order mixing
angles for the PMNS matrix are chosen as special values which can
acquire theoretical backgrounds. With the chosen zeroth value, the
zeroth PMNS matrix is known as several mixing patterns, e.g.,
bimaximal and tribimaximal. The zeroth-order approximation of the
CKM matrix is the unit matrix with all zeroth-order mixing angles
equal to zero. Notice that with the direct application of the QLC
to the zeroth mixing angles, one can find that the zeroth
approximation of the two mixing matrices have a correspondence:
the unit matrix for the CKM matrix corresponds to a bimaximal
pattern for the PMNS matrix~\cite{Li:2005ir}, and the tribimaximal
pattern for the PMNS matrix corresponds to a $V_0$ for the CKM
matrix~\cite{Li:2008aa}, which has the form
\begin{eqnarray}
V_0=\left(
\begin{array}{ccc}
\frac{\sqrt{2}+1}{\sqrt{6}}&\frac{\sqrt{2}-1}{\sqrt{6}}&0\\
-\frac{\sqrt{2}-1}{\sqrt{6}}&\frac{\sqrt{2}+1}{\sqrt{6}}&0\\
0&0&1\\
\end{array}
\right).
\end{eqnarray}
The $V_0$ is a result of the the unified parametrization under the
QLC relation with the tribimaximal pattern as the zeroth for the PMNS
matrix in Ref.~\cite{Li:2008aa}. The triminimal expansion for the
CKM taking $V_0$ as the zeroth-order approximation is discussed in
Ref.~\cite{He:2008td}. The tribimaximal pattern is a good
approximation for the neutrino data before the discovery of large
$\theta_{13}$; consequently, its QLC corespondent $V_0$ possesses the
merit of fast convergence with comparison to the triminimal
expansion based on the unit matrix for the CKM matrix, which is
also pointed out in Refs.~\cite{He:2008td,He:2009jm,Qin:2011ub}.

The triminimal parametrization for the two mixing matrices have been
accomplished with the introduction of the QLC~\cite{He:2009jm}. The main
results in the unified triminimal parametrization are summarized as follows:
\begin{enumerate}
\item The zeroth-order approximation or the base matrix correspondence, i.e., the unit matrix, corresponds to the bimaximal pattern, and the $V_0$ corresponds to the  tribimaximal pattern.
\item The triminimal expansion around $V_0$ converges faster than around the unit matrix.
\end{enumerate}

Though the triminimal parametrization has many advantages, there are three expanding parameters, corresponding to each mixing angle.
In the Wolfenstein parametrization the expansion is measured in
magnitude with only one parameter $\lambda$. This feature, combined
with the motivation to reveal the hierarchical structure, makes it
always beneficial to look for a proper Wolfenstein(-like)
parametrization for the leptons. The work can be done with or without
the QLC relation. We discuss the situation related to the QLC here.

The Wolfenstein parametrization is proposed by reparametrizing the original Kobayashi-Maskawa scheme~\cite{CKM}- which is also one of the angle-phase parametrizations-as the standard one. The Wolfenstein parameters are related to the standard parametrization by the following relations:
\begin{eqnarray}
&&\sin\theta_{12}=\lambda,\nonumber\\
&&\sin\theta_{23}=A \lambda^2,\nonumber\\
&&\sin\theta_{13}e^{i \delta}=A \lambda^3(\rho+i \eta).
\end{eqnarray}
Using these and the QLC relation, the PMNS matrix can be
parametrized by the same Wolfenstein parameters $\lambda$ and $A$ as
in Ref.~\cite{Li:2005ir}. Thus the deviations of the CKM matrix to
its leading order (the unit matrix) and the PMNS matrix to its leading
order (the bimaximal pattern) are all measured by the same expanding
parameter $\lambda$. As there are four mixing parameters in the
Dirac-type neutrino mixing, two additional free parameters are
introduced to accommodate the effect of $\theta_{13}$ and the CP-violating phase $\delta$. Given that not enough information was available on the
smallest mixing angle $\theta_{13}$, some relations of $\theta_{13}$
and $\delta$ with the Wolfenstein parameters $\lambda$ and $A$ were
estimated in previous parametrizations~\cite{Li:2005ir,He:2009jm}, but were adjusted with the observation of a large neutrino mixing angle
$\theta_{13}$~\cite{Zheng:2011uz}. The arbitrariness with some of the degrees
of freedom in choosing the expansion powers will be solved when
other relations are introduced, which we will cover in
Sec.~\ref{sec:correlation}.

The Wolfenstein-like parametrization for the PMNS matrix can also be
done given that $V_0$ is the leading order for the CKM matrix, as in
Ref.~\cite{Li:2008aa}, where the parameters $\lambda$ and $A$ are
introduced according to the magnitude of the matrix elements and are
therefore not necessarily the same as the Wolfenstein parameters and
can be adjusted with the experimental progress in the lepton mixing.

As is pointed out in Ref.~\cite{raidal} and the expansion findings
mentioned above, the CKM matrix can be viewed as describing the
deviation of the PMNS matrix from the bimaximal mixing pattern. This
leads to a conjecture on the correlation of the mixing matrices
themselves, which can be viewed as the matrix form of the QLC.
Explicitly, it can be in the form $V_{\rm CKM}U_{\rm PMNS}=U_{\rm
bimaximal}$ or $U_{\rm PMNS}V_{\rm CKM}=U_{\rm bimaximal}$, as
discussed in Ref.~\cite{Li:2005yj}, where the PMNS matrix is
parametrized with the Wolfenstein parameters with the help of the QLC,
and the product of the PMNS matrix and the CKM matrix are expanded
with $\lambda$, with a result that the bimaximal pattern is the leading
order. It was also pointed out that the QLC relation is maintained at
different orders of $\lambda$, as was discussed in
Ref.~\cite{Zheng:2011uz}. It can also take the form $V_{\rm
CKM}U_{\rm PMNS}=U_{\rm tribimaximal}$ or $U_{\rm PMNS}V_{\rm
CKM}=U_{\rm tribimaximal}$: the corresponding discussion can be found in
Ref.~\cite{Zheng:2011uz}. As the mixing parameters in the CKM matrix
are better determined compared with the PMNS matrix, it is seen as
a good way to parametrize the PMNS matrix with the CKM parameters,
as is done when using the original QLC relation. Another method in this direction
takes the form of the correlation $V_{\rm CKM}^\dagger U_{\rm
PMNS}V_{\rm CKM}=U_{\rm tribimaximal}$~\cite{Qin:2011bq}. Taking the CKM
matrix in the Wolfenstein parametrization and the PMNS matrix in the
standard parametrization, the mixing angles are expressed by the
Wolfenstein parameters; thus, the oscillation probability is expressed
by the Wolfenstein parameters.

\subsection{The self-complementarity (SC) in the lepton mixing}

The self-complementarity (SC) is observed as a phenomenological
relation after the recent results on the relatively large
$\theta_{13}$ in the neutrino mixing. It was suggested in
Ref.~\cite{Zheng:2011uz} when the result of T2K was
released~\cite{T2K}, and systematically studied in
Ref.~\cite{Zhang:2012xu}.

Generally there are three kinds of self-complementarity relations
appearing in the nine angle-phase
parameterizations~\cite{Zhang:2012xu}. Given the discussion
constrained in the standard parametrization, the
self-complementarity relation reads $\theta_{12}^{\rm
PMNS}+\theta_{13}^{\rm PMNS}\simeq\theta_{23}^{\rm PMNS}$. The
verification of the validity of this relation was carried out in the previous subsection.

As the latest experimental results have established the relatively large
$\theta_{13}$, the mixing patterns for the PMNS matrix all face the challenge of accommodating for this result, for they all have a
vanishing $\theta_{13}$. Naively applying the self-complementarity
relation to the mixing patterns to get corrections in $\theta_{13}$,
we find that the tribimaximal requires a rather sound value,
$\theta_{13}=9.736^\circ$~\cite{Zhang:2012xu}. Whether this can
acquire theoretical support still needs to be tested by further
precise measurements.

It is obvious that the self-complementarity offers the correlations
of the lepton mixing angles. Given the correlation inspired by the
self-complementarity, one can find a new mixing pattern for the PMNS
matrix, as in Ref.~\cite{Zheng:2011uz}. By counting the number of
free parameters, it is necessary to make other assumptions to get a
constant matrix as a mixing pattern with special values for the
mixing angles. One such example is reported in
Ref.~\cite{Zheng:2011uz}, where the corresponding triminimal
expansion is also deduced.

Note that once again the freedom of choosing parameters in
determining a leading order (or a mixing pattern) arises. This is
because neither of the relations, i.e., the QLC or the SC, can
sufficiently correlate all the lepton mixing angles.

The self-complementarity relation cannot be deduced in the context
of the standard model. Whether it can acquire some theoretical
support is still an open question. Examining its phenomenological
implications would be helpful before seriously considering this relation from some theoretical backgrounds.

\

\subsection{The new relation $\theta^{\rm PMNS}_{13}=\theta_C/{\sqrt{2}}$}

Based on the latest value of $\theta_{13}^{\rm PMNS}$, an
interesting relation emerges, i.e., $\sin\theta^{\rm
PMNS}_{13}=\frac{1}{\sqrt{2}}\sin\theta_C$, or to a good
approximation, $\theta^{\rm PMNS}_{13}=\frac{1}{\sqrt{2}}\theta_C$.
Numerically, $\frac{1}{\sqrt{2}}\theta_C=(9.21\pm0.04)^\circ$. This
offers another correlation between the CKM parameter and the PMNS
parameter. Actually, the relation $\sin\theta^{\rm
PMNS}_{13}=\frac{1}{\sqrt{2}}\lambda$ shows up as an option from the
QLC in the matrix form as in
Refs.~\cite{qlc,Li:2005yj,Ahn:2011ep,Qin:2011bq}. This relation is discussed
when confronting earlier theoretical predictions with the large
observed $\theta^{\rm PMNS}_{13}$ in Ref.~\cite{Zheng:2011uz}. It is
also suggested as a useful correlation between the CKM and the PMNS
expansion parameters in a new strategy to parametrize the PMNS
matrix in terms of the mixing angle $\theta^{\rm
PMNS}_{13}$~\cite{Ma:2012zm}.

There is a correspondence between the CKM matrix and the PMNS matrix having the form
\begin{eqnarray}
V_M=V_{\rm CKM}\Omega U_{\rm PMNS},
\end{eqnarray}
where $\Omega$ represents the necessary phase between the two mixing matrices when they are in the same representation of a certain gauge group~\cite{Chauhan:2006im}. Leaving the correlation matrix $V_M$ unrestrained and parametrizing the CKM matrix in the Wolfenstein form, we have
\begin{widetext}
\begin{eqnarray}
U_{\rm PMNS}&=&(V_{\rm CKM}\Omega)^\dagger V_M \nonumber\\
&=&\left(
\begin{array}{ccc}
e^{-i\omega_1}&0&0\\
0&e^{-i\omega_2}&0\\
0&0&e^{-i\omega_3}\\
\end{array}
  \right)
\left(
  \begin{array}{ccc}
  1-\frac{1}{2}\lambda^2    & -\lambda                   & A \lambda^3(1-\rho-i\eta) \\
  \lambda                  & 1-\frac{1}{2}\lambda^2    & -A \lambda^2 \\
  A \lambda^3(\rho-i\eta) & A \lambda^2              & 1\\
  \end{array}
  \right)\left(
  \begin{array}{ccc}
  V_{11}&V_{12}&V_{13}\\
  V_{21}&V_{22}&V_{23}\\
  V_{31}&V_{32}&V_{33}\\
  \end{array}
 \right)
+\mathcal{O}(\lambda^4).\label{Vm}
\end{eqnarray}
\end{widetext}
Consequently, we have
\begin{eqnarray}
U_{e3}=\left((1-\frac{\lambda^2}{2})V_{13}-\lambda V_{23}+A \lambda^3 V_{33}(1-\rho-i\eta)\right)e^{-i \omega_1}.~~
\end{eqnarray}
For $V_{13}=0$ and $V_{23}=\frac{1}{\sqrt{2}}$, which is just the situation for $V_M=U_{\rm bimaximal}$ and $V_M=U_{\rm tribimaximal}$, we have $U_{e3}=\frac{1}{\sqrt{2}}\lambda$ in $\mathcal{O}(\lambda^2)$. Notice that $|U_{e3}|=\sin\theta_{13}$ in the standard parametrization, so we get the relation under discussion.

We would like to add a few comments here on the differences of the $U_{\rm PMNS}$ and $V_{\rm CKM}$. By introducing the right-handed neutrinos into the framework of the standard model, we have a Dirac mass term for neutrinos which is the same as that for other fermions. When expressing the charged current interaction in the basis of definite mass, the mixing matrices are introduced. It is known that such a direct way of introducing the neutrino mass shares the most similarities with quarks and leptons, but it cannot account for the smallness of the neutrino mass. The seesaw formula offers a natural explanation for the origin of the light neutrino mass by introducing heavy Majorana neutrinos. We have the mass matrix of the light neutrinos in the form
\begin{eqnarray}
\mathcal{M_{\rm seesaw}}=\mathcal{M_{\rm Dirac}}\frac{1}{\mathcal{M_{\rm Majorana}}}\mathcal{M_{\rm Dirac}^{\rm T}},
\end{eqnarray}
where $\mathcal{M_{\rm Dirac}}$ corresponds to the Yukawa coupling of the right-handed and left-handed neutrinos. The Dirac mass term can be diagonalized by two unitary matrices which rotates the left-handed and right-handed fields separately. $\mathcal{M_{\rm seesaw}}$ is diagonalized as
\begin{eqnarray}
\mathcal{M}_{\rm seesaw}=U_{\rm L}^{\nu} \mathcal{F} \mathcal{M}_{\rm seesaw}^{\rm diag} \mathcal{F}^{\rm T} U_{\rm L}^{\nu T},
\end{eqnarray}
where $U_{\rm L}^{\nu}$ represents the unitary matrix that rotates the left-handed fields as in the Dirac mass term. The resulting $U_{\rm PMNS}$ is of the form
\begin{eqnarray}
U_{\rm PMNS}=U_{\rm L}^{l \dagger} U_{\rm L}^{\nu} \mathcal{F},\label{Uss}
\end{eqnarray}
where $\mathcal{F}$ shows explicitly the difference between the $U_{\rm PMNS}$ and $V_{\rm CKM}$ matrices.

After using the correlations of the mass matrices (or equivalently, the Yukawa coupling matrices) in the SU(5) and SO(10) GUTs, Eq.~(\ref{Uss}) takes the form,
\begin{eqnarray}
U_{\rm PMNS}=V_{\rm CKM}^{\dagger} \mathcal{F},
\end{eqnarray}
which is the same form as in Eq.~(\ref{Vm})~\cite{Datta:2005ci}.

It is worth mentioning that the above discussion is based on the
Type \uppercase\expandafter{\romannumeral1} seesaw mechanism.
Whether the results can be achieved in other frameworks of mass
generation, e.g., a double seesaw mechanism, remains an open question.

King noticed this relation~\cite{King:2012vj} and proposed a
tribimaximal-Cabbibo (TBC) mixing with
$\sin\theta_{13}=\sin\theta_C/\sqrt{2},\quad\sin\theta_{23}=1/\sqrt{2},\quad\sin\theta_{13}=1/\sqrt{3}$.
Its realization in the Type \uppercase\expandafter{\romannumeral1}
seesaw mechanism is also discussed. The realization of $\theta^{\rm
PMNS}_{13}=\frac{1}{\sqrt{2}}\theta_C$ via charged lepton
corrections in SU(5) GUTs and Pati-Salam models is discussed in
Ref.~\cite{Antusch:2012fb}. $U_{e3}=\theta_C/\sqrt{2}$ could also
happen as an option from some consideration of flavor
symmetry~\cite{Masina}.

\section{\label{sec:correlation}CORRELATIONS OF DIFFERENT SCENARIOS AND CORRESPONDING PMNS PARAMETRIZATIONS}
There are four relations under discussion. As there are three lepton
mixing angles, all four relations are not expected to be
simultaneously true. By taking three of the four relations to be simultaneously
valid, we find that they can form four scenarios. In each scenario
the three lepton mixing angles are related to the Wolfenstein
parameters in the quark mixing in a unique way. Consequently, the
correlation for the lepton mixing angles in each scenario is also
unique.

Firstly, we restate the four relations for later use and convenience:
\begin{eqnarray}
&&\theta_{12}^{\rm CKM}+\theta_{12}^{\rm PMNS}=45^\circ,\label{qlc1}\\
&&\theta_{23}^{\rm CKM}+\theta_{23}^{\rm PMNS}=45^\circ,\label{qlc2}\\
&&\theta_{12}^{\rm PMNS}+\theta_{13}^{\rm PMNS}=\theta_{23}^{\rm PMNS},\label{sc}\\
&&\sin\theta_{13}^{\rm PMNS}=\frac{\lambda}{\sqrt{2}}.\label{c}
\end{eqnarray}

The four scenarios are as follows:
\begin{enumerate}
\item Scenario A\\
Taking relations in Eq.~(\ref{qlc1}), Eq.~(\ref{sc}), and Eq.~(\ref{c}) to be simultaneously true.
\item Scenario B\\
Taking relations in Eq.~(\ref{qlc2}), Eq.~(\ref{sc}), and Eq.~(\ref{c}) to be simultaneously true.
\item Scenario C\\
Taking relations in Eq.~(\ref{qlc1}), Eq.~(\ref{qlc2}), and Eq.~(\ref{c}) to be simultaneously true.
\item Scenario D\\
Taking relations in Eq.~(\ref{qlc1}), Eq.~(\ref{qlc2}), and Eq.~(\ref{sc}) to be simultaneously true.
\end{enumerate}

Next we discuss each scenario and give the explicit form of the
correlations as well as the corresponding PMNS parametrizations.

\subsection{Scenario A}

By taking relations in Eq.~(\ref{qlc1}), Eq.~(\ref{sc}), and
Eq.~(\ref{c}) to be simultaneously true, namely,
\begin{eqnarray}
&&\theta_{12}^{\rm CKM}+\theta_{12}^{\rm PMNS}=45^\circ,\nonumber\\
&&\theta_{12}^{\rm PMNS}+\theta_{13}^{\rm PMNS}=\theta_{23}^{\rm PMNS},\nonumber\\
&&\sin\theta_{13}^{\rm PMNS}=\frac{\lambda}{\sqrt{2}},\nonumber
\end{eqnarray}
we find that the trigonometric functions of the mixing angles in the PMNS matrix can be expressed in the following way, to $\mathcal{O}(\lambda^4)$:
\begin{eqnarray}
\sin\theta_{13}^{\rm PMNS}&=&\frac{\lambda}{\sqrt{2}},\\
\cos\theta_{13}^{\rm PMNS}&=&1-\frac{\lambda^2}{4}+\mathcal{O}(\lambda^4),\\
\sin\theta_{12}^{\rm PMNS}&=&\frac{1}{\sqrt{2}}(1-\lambda-\frac{\lambda^2}{2})+\mathcal{O}(\lambda^4),\\
\cos\theta_{12}^{\rm PMNS}&=&\frac{1}{\sqrt{2}}+\frac{\lambda}{\sqrt{2}}-\frac{\lambda^2}{2\sqrt{2}}+\mathcal{O}(\lambda^4),\\
\sin\theta_{23}^{\rm PMNS}&=&\frac{1}{\sqrt{2}}+(\frac{1}{2}-\frac{1}{\sqrt{2}})\lambda+
(\frac{1}{2}-\frac{3}{4\sqrt{2}})\lambda^2\nonumber\\
&&+\frac{1}{8}(-2+\sqrt{2})\lambda^3+\mathcal{O}(\lambda^4),\\
\cos\theta_{23}^{\rm PMNS}&=&\frac{1}{\sqrt{2}}+(-\frac{1}{2}+\frac{1}{\sqrt{2}})\lambda+
(\frac{1}{2}-\frac{3}{4\sqrt{2}})\lambda^2\nonumber\\
&&+\frac{1}{8}(2-\sqrt{2})\lambda^3+\mathcal{O}(\lambda^4).
\end{eqnarray}
Substituting the trigonometric functions with the corresponding
expansion in $\lambda$ into the standard parametrization, we have
\begin{widetext}
\begin{eqnarray}
U_{\rm PMNS}&=&\left(
  \begin{array}{ccc}
    \frac{1}{\sqrt{2}}  & \frac{1}{\sqrt{2}}     & 0         \\
    -\frac{1}{2}        & \frac{1}{2}   & \frac{1}{\sqrt{2}} \\
    \frac{1}{2}         & -\frac{1}{2}  & \frac{1}{\sqrt{2}} \\
  \end{array}
\right) \nonumber
+\lambda\left(
  \begin{array}{ccc}
    \frac{1}{\sqrt{2}}             & -\frac{1}{\sqrt{2}} & \frac{e^{-i\delta}}{\sqrt{2}} \\
    \frac{1}{2\sqrt{2}}-\frac{e^{-i\delta}}{2\sqrt{2}}&
    1-\frac{1}{2\sqrt{2}}-\frac{e^{-i\delta}}{2\sqrt{2}} & \frac{1}{2}-\frac{1}{\sqrt{2}}\\      -1+\frac{1}{2\sqrt{2}}-\frac{e^{-i\delta}}{2\sqrt{2}}&
     -\frac{1}{2\sqrt{2}}-\frac{e^{-i\delta}}{2\sqrt{2}}& -\frac{1}{2}+\frac{1}{\sqrt{2}}\\
  \end{array}
\right) \nonumber\\
&&+\lambda^2\left(
  \begin{array}{ccc}
  -\frac{3}{4\sqrt{2}}& -\frac{3}{4\sqrt{2}}& 0\\
  \frac{9}{8}-\frac{1}{\sqrt{2}}-\frac{e^{i\delta}}{4}&
  -\frac{1}{8}-\frac{e^{i\delta}}{4}+\frac{e^{i\delta}}{\sqrt{2}}&
  \frac{1}{2}-\frac{1}{\sqrt{2}}\\
   -\frac{1}{8}+\frac{e^{i\delta}}{4}-\frac{e^{i\delta}}{\sqrt{2}}&
   \frac{9}{8}-\frac{1}{\sqrt{2}}+\frac{e^{i\delta}}{4}&
   \frac{1}{2}-\frac{1}{\sqrt{2}}\\
    \end{array}
\right)\nonumber\\
&&+\lambda^3\left(
  \begin{array}{ccc}
  -\frac{1}{4\sqrt{2}}&\frac{1}{4\sqrt{2}}&0\\
  -\frac{e^{i\delta}}{2}+\frac{9e^{i\delta}}{8\sqrt{2}}&
  -\frac{3}{4}+\frac{1}{\sqrt{2}}+\frac{e^{i\delta}}{8\sqrt{2}}&
  -\frac{3}{8}+\frac{1}{2\sqrt{2}}\\
  \frac{3}{4}-\frac{1}{\sqrt{2}}+\frac{e^{i\delta}}{8\sqrt{2}}&
  -\frac{e^{i\delta}}{2}+\frac{9e^{i\delta}}{8\sqrt{2}}&
  \frac{3}{8}-\frac{1}{2\sqrt{2}}\\
  \end{array}
\right)
 +\mathcal{O}(\lambda^4).
\end{eqnarray}
\end{widetext}
We add a few comments here.
\begin{enumerate}
\item We do not use Eq.~(\ref{qlc2}) because it relates the $\theta_{23}^{\rm PMNS}$ to $\theta_{23}^{\rm CKM}$, where $\sin\theta_{23}^{\rm CKM}=A \lambda^2$. In this scenario, $\theta_{23}^{\rm PMNS}$ relates to the Wolfenstein parameter $\lambda$ indirectly from $\theta_{12}^{\rm PMNS}$ and $\theta_{13}^{\rm PMNS}$.
\item A simple check for Eq.~(\ref{qlc2}) is $\sin(\theta_{23}^{\rm CKM}+\theta_{23}^{\rm PMNS})=\frac{1}{\sqrt{2}}+(\frac{1}{2}-\frac{1}{\sqrt{2}})\lambda+\mathcal{O}(\lambda^2)$, which means that the four relations cannot be simultaneously exact. Taking relations in Eq.~(\ref{qlc1}), Eq.~(\ref{sc}), and Eq.~(\ref{c}) to be simultaneously true results in a correction in Eq.~(\ref{qlc2}).
\item The bimaximal pattern is achieved naturally in the leading order, which is not so obvious, as we admit only one QLC relation here.
\item Eq.~(\ref{c}) is arrived at $\mathcal{O}(\lambda)$, which is not surprising as it is one of the conditions.
\item As a useful quantity for the measurement of the CP violation, we calculate the Jarlskog invariant in this scenario, which is $J=\frac{1}{4 \sqrt{2}}\sin\delta\lambda +\left(\frac{1}{2}-\frac{11 }{8 \sqrt{2}}\right)\sin\delta \lambda ^3+\mathcal{O}(\lambda^5)=0.0346\sin\delta$.
\item The simplified Wolfenstein parametrization for the PMNS matrix in this scenario has only two parameters: the same $\lambda$ as in the quark mixing and a CP violating phase $\delta$.
\item The PMNS matrix in this Wolfenstein parametrization requires corrections in the size of $\lambda$, which is also the expanding parameter in the CKM matrix.
\end{enumerate}

\subsection{Scenario B}

By taking relations in Eq.~(\ref{qlc2}), Eq.~(\ref{sc}), and
Eq.~(\ref{c}), i.e.,
\begin{eqnarray}
&&\theta_{23}^{\rm CKM}+\theta_{23}^{\rm PMNS}=45^\circ,\nonumber\\
&&\theta_{12}^{\rm PMNS}+\theta_{13}^{\rm PMNS}=\theta_{23}^{\rm PMNS},\nonumber\\
&&\sin\theta_{13}^{\rm PMNS}=\frac{\lambda}{\sqrt{2}},\nonumber
\end{eqnarray}
to be simultaneously true, following a similar procedure as in the previous subsection, we have
\begin{eqnarray}
\sin\theta_{13}^{\rm PMNS}&=&\frac{\lambda}{\sqrt{2}},\\
\cos\theta_{13}^{\rm PMNS}&=&1-\frac{\lambda^2}{4}+\mathcal{O}(\lambda^4),\\
\sin\theta_{12}^{\rm PMNS}&=&\frac{1}{\sqrt{2}}-\frac{\lambda }{2}+\left(-\frac{1}{4 \sqrt{2}}-\frac{A}{\sqrt{2}}\right) \lambda ^2\nonumber\\
&&-\frac{A \lambda ^3}{2}+\mathcal{O}(\lambda^4),\\
\cos\theta_{12}^{\rm PMNS}&=&\frac{1}{\sqrt{2}}+\frac{\lambda }{2}+\left(-\frac{1}{4 \sqrt{2}}+\frac{A}{\sqrt{2}}\right) \lambda ^2\nonumber\\
&&-\frac{A \lambda ^3}{2}+\mathcal{O}(\lambda^4),\\
\sin\theta_{23}^{\rm PMNS}&=&\frac{1}{\sqrt{2}}-\frac{A \lambda ^2}{\sqrt{2}}+\mathcal{O}(\lambda^4),\\
\cos\theta_{23}^{\rm PMNS}&=&\frac{1}{\sqrt{2}}+\frac{A \lambda ^2}{\sqrt{2}}+\mathcal{O}(\lambda^4).
\end{eqnarray}
Substituting the trigonometric functions with the corresponding
expansion in $\lambda$ into the standard parametrization, we have
\begin{widetext}
\begin{eqnarray}
U_{\rm PMNS}&=&\left(
  \begin{array}{ccc}
    \frac{1}{\sqrt{2}}  & \frac{1}{\sqrt{2}}     & 0         \\
    -\frac{1}{2}        & \frac{1}{2}   & \frac{1}{\sqrt{2}} \\
    \frac{1}{2}         & -\frac{1}{2}  & \frac{1}{\sqrt{2}} \\
  \end{array}
\right) \nonumber
+\lambda\left(
  \begin{array}{ccc}
    \frac{1}{2}                     & -\frac{1}{2}                    & \frac{e^{-i\delta}}{\sqrt{2}} \\
    \frac{-1+e^{i\delta}}{2\sqrt{2}}&-\frac{-1+e^{i\delta}}{2\sqrt{2}}&0\\
    -\frac{1+e^{i\delta}}{2\sqrt{2}}&-\frac{1+e^{i\delta}}{2\sqrt{2}} &0\\
  \end{array}
\right) \nonumber\\
&&+\lambda^2\left(
  \begin{array}{ccc}
  -\frac{1}{2\sqrt{2}}+\frac{A}{\sqrt{2}}& -\frac{1}{2\sqrt{2}}-\frac{A}{\sqrt{2}}&0\\
  \frac{1}{8}-\frac{e^{i\delta}}{4} &-\frac{1}{8}+A+\frac{e^{i\delta}}{4}&-\frac{1}{4\sqrt{2}}-\frac{A}{\sqrt{2}}\\
  -\frac{1}{8}-A-\frac{e^{i\delta}}{4}& \frac{1}{8}+\frac{e^{i\delta}}{4}& -\frac{1}{4\sqrt{2}}+\frac{A}{\sqrt{2}}\\
    \end{array}
\right)\nonumber\\
&&+\lambda^3\left(
  \begin{array}{ccc}
  -\frac{1}{8}-\frac{A}{2}            &\frac{1}{8}-\frac{A}{2}            &0\\
  \frac{8A+e^{i\delta}}{8\sqrt{2}}    &\frac{(1+8A)e^{i\delta}}{8\sqrt{2}}&0\\
  \frac{(-1+8A)e^{i\delta}}{8\sqrt{2}}&\frac{8A+e^{i\delta}}{8\sqrt{2}}   &0\\
  \end{array}
\right)
 +\mathcal{O}(\lambda^4).
\end{eqnarray}
\end{widetext}

\begin{enumerate}
\item A simple check for Eq.~(\ref{qlc1}) is $\sin(\theta_{12}^{\rm CKM}+\theta_{12}^{\rm PMNS})=\frac{1}{\sqrt{2}}+(-\frac{1}{2}+\frac{1}{\sqrt{2}})\lambda+\mathcal{O}(\lambda^2)$.
\item The bimaximal pattern is achieved in the leading order again, and also only one QLC relation is taken here.
\item Eq.~(\ref{c}) is arrived at $\mathcal{O}(\lambda)$.
\item The Jarlskog invariant in this scenario is $J=\frac{1}{4 \sqrt{2}}\sin\delta\lambda-\frac{3 }{8 \sqrt{2}} \sin\delta\lambda ^3-\frac{A}{2}  \sin\delta \lambda ^4+\mathcal{O}(\lambda^5)=0.0356\sin\delta$.
\item The simplified Wolfenstein parametrization for the PMNS matrix in this scenario has three parameters: the same $A$ and $\lambda$ as in the quark mixing, and a CP-violating phase $\delta$.
\item The PMNS matrix in this Wolfenstein parametrization requires corrections in the size of $\lambda$, which is also the expanding parameter in the CKM matrix.
\end{enumerate}

\subsection{Scenario C}

By taking relations in Eq.~(\ref{qlc1}), Eq.~(\ref{qlc2}), and
Eq.~(\ref{c}), i.e.,
\begin{eqnarray}
&&\theta_{12}^{\rm CKM}+\theta_{12}^{\rm PMNS}=45^\circ,\nonumber\\
&&\theta_{23}^{\rm CKM}+\theta_{23}^{\rm PMNS}=45^\circ,\nonumber\\
&&\sin\theta_{13}^{\rm PMNS}=\frac{\lambda}{\sqrt{2}},\nonumber
\end{eqnarray}
to be simultaneously true, we have
\begin{eqnarray}
\sin\theta_{13}^{\rm PMNS}&=&\frac{\lambda}{\sqrt{2}},\\
\cos\theta_{13}^{\rm PMNS}&=&1-\frac{\lambda^2}{4}+\mathcal{O}(\lambda^4),\\
\sin\theta_{12}^{\rm PMNS}&=&\frac{1}{\sqrt{2}}-\frac{\lambda }{\sqrt{2}}-\frac{\lambda ^2}{2 \sqrt{2}}+\mathcal{O}(\lambda^4),\\
\cos\theta_{12}^{\rm PMNS}&=&\frac{1}{\sqrt{2}}+\frac{\lambda }{\sqrt{2}}-\frac{\lambda ^2}{2 \sqrt{2}}+\mathcal{O}(\lambda^4),\\
\sin\theta_{23}^{\rm PMNS}&=&\frac{1}{\sqrt{2}}-\frac{A \lambda ^2}{\sqrt{2}}+\mathcal{O}(\lambda^4),\\
\cos\theta_{23}^{\rm PMNS}&=&\frac{1}{\sqrt{2}}+\frac{A \lambda ^2}{\sqrt{2}}+\mathcal{O}(\lambda^4).
\end{eqnarray}
Substituting the trigonometric functions with the corresponding
expansion in $\lambda$ into the standard parametrization, we have
\begin{widetext}
\begin{eqnarray}
U_{\rm PMNS}&=&\left(
  \begin{array}{ccc}
    \frac{1}{\sqrt{2}}  & \frac{1}{\sqrt{2}}     & 0         \\
    -\frac{1}{2}        & \frac{1}{2}   & \frac{1}{\sqrt{2}} \\
    \frac{1}{2}         & -\frac{1}{2}  & \frac{1}{\sqrt{2}} \\
  \end{array}
\right) \nonumber
+\lambda\left(
  \begin{array}{ccc}
    \frac{1}{\sqrt{2}}             & -\frac{1}{\sqrt{2}} & \frac{e^{-i\delta}}{\sqrt{2}} \\
    \frac{1}{2}-\frac{e^{i\delta}}{2\sqrt{2}} &\frac{1}{2}-\frac{e^{i\delta}}{2\sqrt{2}} &0\\
    -\frac{1}{2}-\frac{e^{i\delta}}{2\sqrt{2}}&-\frac{1}{2}-\frac{e^{i\delta}}{2\sqrt{2}}&0\\
  \end{array}
\right) \nonumber\\
&&+\lambda^2\left(
  \begin{array}{ccc}
  -\frac{3}{4\sqrt{2}}& -\frac{3}{4\sqrt{2}}& 0\\
  \frac{1}{4}(1-2A-\sqrt{2}e^{i \delta})&\frac{1}{4}(-1+2A+\sqrt{2}e^{i \delta})&-\frac{1}{4\sqrt{2}}-\frac{A}{\sqrt{2}}\\
  \frac{1}{4}(-1-2A-\sqrt{2}e^{i \delta})&\frac{1}{4}(1+2A+\sqrt{2}e^{i \delta})&-\frac{1}{4\sqrt{2}}+\frac{A}{\sqrt{2}}\\
    \end{array}
\right)\nonumber\\
&&+\lambda^3\left(
  \begin{array}{ccc}
  -\frac{1}{4\sqrt{2}}&\frac{1}{4\sqrt{2}}&0\\
  \frac{1}{8}(4A+\sqrt{2}e^{i\delta}+2\sqrt{2}Ae^{i\delta})&\frac{1}{8}(4A+\sqrt{2}e^{i\delta}+2\sqrt{2}Ae^{i\delta})&0\\
  \frac{1}{8}(4A+\sqrt{2}e^{i\delta}-2\sqrt{2}Ae^{i\delta})&\frac{1}{8}(4A+\sqrt{2}e^{i\delta}-2\sqrt{2}Ae^{i\delta})&0\\
  \end{array}
\right)
 +\mathcal{O}(\lambda^4).
\end{eqnarray}
\end{widetext}

\begin{enumerate}
\item A check for Eq.~(\ref{sc}) is $\sin(\theta_{12}^{\rm PMNS}+\theta_{13}^{\rm PMNS})-\sin\theta_{23}^{\rm PMNS}=(\frac{1}{2}-\frac{1}{\sqrt{2}})\lambda+\mathcal{O}(\lambda^2)$.
\item The bimaximal pattern is achieved in the leading order, which is a direct result of two QLC relations.
\item Eq.~(\ref{c}) is arrived at $\mathcal{O}(\lambda)$.
\item The Jarlskog invariant in this scenario is $J=\frac{1}{4\sqrt{2}}\sin\delta\lambda -\frac{5}{8\sqrt{2}}\sin\delta\lambda^3+\mathcal{O}(\lambda^5)=0.0348\sin\delta$.
\item The simplified Wolfenstein parametrization for the PMNS matrix in this scenario also has three parameters: the same $A$ and $\lambda$ as in the quark mixing, and a CP-violating phase $\delta$.
\item The PMNS matrix in this Wolfenstein parametrization requires corrections in the size of $\lambda$, which is also the expanding parameter in the CKM matrix.
\end{enumerate}

\subsection{Scenario D}

By taking relations in Eq.~(\ref{qlc1}), Eq.~(\ref{qlc2}), and
Eq.~(\ref{sc}), i.e.,
\begin{eqnarray}
&&\theta_{12}^{\rm CKM}+\theta_{12}^{\rm PMNS}=45^\circ,\nonumber\\
&&\theta_{23}^{\rm CKM}+\theta_{23}^{\rm PMNS}=45^\circ,\nonumber\\
&&\theta_{12}^{\rm PMNS}+\theta_{13}^{\rm PMNS}=\theta_{23}^{\rm PMNS},\nonumber
\end{eqnarray}
to be simultaneously true, we have
\begin{eqnarray}
\sin\theta_{13}^{\rm PMNS}&=&\lambda -A \lambda ^2+\mathcal{O}(\lambda^4),\\
\cos\theta_{13}^{\rm PMNS}&=&1-\frac{\lambda ^2}{2}+A \lambda ^3+\mathcal{O}(\lambda^4),\\
\sin\theta_{12}^{\rm PMNS}&=&\frac{1}{\sqrt{2}}-\frac{\lambda }{\sqrt{2}}-\frac{\lambda ^2}{2 \sqrt{2}}+\mathcal{O}(\lambda^4),\\
\cos\theta_{12}^{\rm PMNS}&=&\frac{1}{\sqrt{2}}+\frac{\lambda }{\sqrt{2}}-\frac{\lambda ^2}{2 \sqrt{2}}+\mathcal{O}(\lambda^4),\\
\sin\theta_{23}^{\rm PMNS}&=&\frac{1}{\sqrt{2}}-\frac{A \lambda ^2}{\sqrt{2}}+\mathcal{O}(\lambda^4),\\
\cos\theta_{23}^{\rm PMNS}&=&\frac{1}{\sqrt{2}}+\frac{A \lambda ^2}{\sqrt{2}}+\mathcal{O}(\lambda^4).
\end{eqnarray}
Substituting the trigonometric functions with the corresponding
expansion in $\lambda$ into the standard parametrization, we have
\begin{widetext}
\begin{eqnarray}
U_{\rm PMNS}&=&\left(
  \begin{array}{ccc}
    \frac{1}{\sqrt{2}}  & \frac{1}{\sqrt{2}}     & 0         \\
    -\frac{1}{2}        & \frac{1}{2}   & \frac{1}{\sqrt{2}} \\
    \frac{1}{2}         & -\frac{1}{2}  & \frac{1}{\sqrt{2}} \\
  \end{array}
\right) \nonumber
+\lambda\left(
  \begin{array}{ccc}
    \frac{1}{\sqrt{2}}             & -\frac{1}{\sqrt{2}} & e^{-i\delta} \\
    \frac{1}{2}-\frac{e^{i\delta}}{2}&\frac{1}{2}-\frac{e^{i\delta}}{2}&0\\
    -\frac{1}{2}-\frac{e^{i\delta}}{2}&-\frac{1}{2}-\frac{e^{i\delta}}{2}&0\\
  \end{array}
\right) \nonumber\\
&&+\lambda^2\left(
  \begin{array}{ccc}
  -\frac{1}{\sqrt{2}}&-\frac{1}{\sqrt{2}}&-Ae^{-i\delta}\\
  \frac{1}{4}(1-2A-2e^{i\delta}+2Ae^{i\delta})&\frac{1}{4}(-1+2A+2e^{i\delta}+2Ae^{i\delta})&-\frac{1}{2\sqrt{2}}-\frac{A}{\sqrt{2}}\\
  \frac{1}{4}(-1-2A-2e^{i\delta}+2Ae^{i\delta})&\frac{1}{4}(1+2A+2e^{i\delta}+2Ae^{i\delta})&-\frac{1}{2\sqrt{2}}+\frac{A}{\sqrt{2}}\\
    \end{array}
\right)\nonumber\\
&&+\lambda^3\left(
  \begin{array}{ccc}
  -\frac{1}{2\sqrt{2}}+\frac{A}{\sqrt{2}}&-\frac{1}{2\sqrt{2}}+\frac{A}{\sqrt{2}}&0\\
  \frac{1}{4}(2A+e^{i\delta}+4Ae^{i\delta})&\frac{1}{4}(2A+e^{i\delta})&\frac{A}{\sqrt{2}}\\
  \frac{1}{4}(2A+e^{i\delta})&\frac{1}{4}(2A+e^{i\delta}-4Ae^{i\delta})&\frac{A}{\sqrt{2}}\\
  \end{array}
\right)
 +\mathcal{O}(\lambda^4).
\end{eqnarray}
\end{widetext}

\begin{enumerate}
\item A check for Eq.~(\ref{c}) is $\sin\theta_{13}^{\rm PMNS}=\lambda+\mathcal{O}(\lambda^2)$.
\item The bimaximal pattern is achieved in the leading order, which is a direct result of two QLC relations.
\item The Jarlskog invariant in this scenario is $J=\frac{1}{4}\sin\delta\lambda -\frac{A}{4}\sin\delta\lambda^2-\frac{3}{4}\sin\delta\lambda^3+\frac{11A}{8}\sin\delta\lambda^4+\mathcal{O}(\lambda^5)=0.0400\sin\delta$.
\item The simplified Wolfenstein parametrization for the PMNS matrix in this scenario has three parameters: the same $A$ and $\lambda$ as in the quark mixing, and a CP-violating phase $\delta$.
\item The PMNS matrix in this Wolfenstein parametrization requires corrections in the size of $\lambda$, which is also the expanding parameter in the CKM matrix.
\end{enumerate}

It is notable that to $\mathcal{O}(\lambda^2)$, Scenarios A, B and C all require a $7\% $ correction to the relation not being used, while Scenario D requires a $41\%$ correction. This means that Scenario D is not desirable in this sense.

\section{\label{sec:discuss}DISCUSSIONS AND CONCLUSIONS}
There are two questions concerning our motivation here. One is, why are we looking for a Wolfenstein-like parametrization in the lepton mixing? The other is, why is the same $\lambda$ as in the CKM matrix important for the Wolfenstein-like parametrization in the lepton sector?

When answering the first question, one should notice the simple fact that the Wolfenstein(-like) parametrization reveals the hierarchical structure with only one expanding parameter, which distinguishes it from the triminimal approach. Only one expanding parameter means that the limiting behavior can be analyzed more directly. So it naturally leads to the answer to the second question: if we have the same expanding parameter for the CKM and the PMNS matrix, their behavior in the limit of a vanishing expanding parameter can be analyzed in a unified way.

Using $\lambda$ as the single parameter characterizing the mixing in both the quark and lepton sectors can be tracked to Raidal~\cite{raidal}, where a non-Abelian flavor model is presented. The same idea is proposed in Ref.~\cite{Datta:2005ci} as Cabibbo haze in the context of grand
unification, and further explained in
Ref.~\cite{Everett:2005ku}. In those works, it is argued that the lepton mixing acquires Cabibbo-sized effects from an 
unknown base in the $\lambda\rightarrow0$ limit where the haze
originates. Reference~\cite{Everett:2005ku} also states that $\lambda$ could be the order parameter characterizing the flavor structures of the quarks and leptons. We find that the lepton mixing does acquire a
Cabibbo-sized effect given the phenomenological relations, and that the
base is the well-known bimaximal pattern. Therefore, it is no longer a haze
given these phenomenological relations.

With a proper redefinition of the mixing angles in the PMNS matrix, or
more straightforwardly, starting with the magnitude of the matrix
elements, one can parametrize the PMNS matrix in a Wolfenstein-like
way as in
Refs.~\cite{Ma:2012zm,Li:2004nh,Rodejohann:2003sc,Li:2004dn,King:2007pr}.
In Ref.~\cite{Ma:2012zm}, a Wolfenstein-like expansion of the PMNS
matrix is achieved with $\theta_{13}$ as the expanding parameter,
resulting in a parametrization that is simple in form, convenient for
use, and  general with all independent degrees of freedom.

In this work, we started by revisiting the phenomenological
relations concerning fermion mixing angles. Their roles in the
neutrino phenomenology, especially in the direction towards the
quark-lepton unification by certain connections were emphasized. We then
analyzed the unique correlation of the lepton mixing angles given
by the relations in four scenarios. We made the simplified
Wolfenstein parametrization for the PMNS matrix. Thus the
hierarchical structure in the lepton sector is revealed with the
same expanding parameter $\lambda$ as in the quark mixing. We
emphasize that by taking the indication from the phenomenological
relations, the resulting Wolfenstein parametrization is restrictive
without arbitrariness in choosing free parameters by hand, and
simplified in the sense that less parameters show up. This feature
has the benefit of being definitive and the drawback of being restrictive.
It calls for futther experimental progress to verify the
phenomenological relations we have taken as our primary condition. The
phenomenological relations may all acquire corrections when more
restrictive data are available, but this may not hurt the discussion
in this work, as seeking connections may inspire more profound
thoughts than were ever expected.

\begin{acknowledgments}
This work is partially supported by National Natural Science
Foundation of China (Grants No.~11021092, No.~10975003,
No.~11035003, and No.~11120101004) and by the Research Fund for the
Doctoral Program of Higher Education (China).
\end{acknowledgments}


\begin{thebibliography}{99}
\bibitem{CKM}
N.~Cabibbo, Phys.\ Rev.\ Lett.\  {\bf 10},  531 (1963);\\
M.~Kobayashi and T.~Maskawa, Prog.\ Theor.\ Phys.\  {\bf 49}, 652
(1973).

\bibitem{PMNS}
B.~Pontecorvo, Sov.\ Phys.\ JETP {\bf 26},  984 (1968);\\
Z.~Maki, M.~Nakagawa and S.~Sakata, Prog.\ Theor.\ Phys.\  {\bf 28},
870 (1962).

\bibitem{CK}
L.L.~Chau and W.Y.~Keung,
Phys. Rev. Lett. {\bf 53}, 1802 (1984).

\bibitem{PDG1996}
R.M.~Barnett {\it et al.}, [Particle Data Group],  Phys. Rev. D {\bf
54}, 1 (1996).

\bibitem{pdg2010}
K. Nakamura {\it et al.} [Particle Data Group], J. Phys. G {\bf 37},
075021 (2010).

\bibitem{Wolfenstein:1983yz}
  L.~Wolfenstein,
  Phys.\ Rev.\ Lett.\  {\bf 51}, 1945 (1983).

\bibitem{bi}
F.~Vissani, arXiv:hep-ph/9708483; V.D.~Barger, S.Pakvasa, T.J.~Weiler, and
K.~Whisnant, Phys. Lett. B {\bf 437}, 107 (1998); A.J.~Baltz,
A.S.~Goldhaber, and M.~Goldhaber, Phys. Rev. Lett. {\bf 81}, 5730
(1998); I.~Stancu and D.V.~Ahluwalia, Phys. Lett. B {\bf 460}, 431
(1999); H.~Georgi and S.L.~Glashow, Phy. Rev. D {\bf 61}, 097301
(2000);
  N.~Li and B.-Q.~Ma,
  Phys.\ Lett.\  B {\bf 600}, 248 (2004)
  [arXiv:hep-ph/0408235].


\bibitem{tri}
P.F.~Harrison, D.H.~Perkins, and W.G.~Scott, Phys. Lett. B {\bf 458},
79 (1999); Phys. Lett. B {\bf 530}, 167 (2002); Z.Z.~Xing, Phys.
Lett. B {\bf 533}, 85 (2002); P.F.~Harrison and W.G.~Scott, Phys.
Lett. B {\bf 535}, 163 (2002); Phys. Lett. B {\bf 557}, 76 (2003);
X.-G.~He and A.~Zee, Phys. Lett. B {\bf 560}, 87 (2003);
  N.~Li and B.-Q.~Ma,
  Phys.\ Rev.\  D {\bf 71}, 017302 (2005)
  [arXiv:hep-ph/0412126].
See also L.~Wolfenstein, Phys. Rev. D {\bf 18}, 958 (1978);
Y.~Yamanaka, H.~Sugawara, and S.~Pakvasa, Phys. Rev. D {\bf 25}, 1895
(1982); D {\bf 29}, 2135(E) (1984).




\bibitem{T2K}
  K.~Abe {\it et al.}  [T2K Collaboration],
  Phys.\ Rev.\ Lett.\  {\bf 107}, 041801 (2011).
\bibitem{expt2}
  P.~Adamson {\it et al.}  [MINOS Collaboration],
  Phys.\ Rev.\ Lett.\  {\bf 107}, 181802 (2011).
\bibitem{expt3}
  Y.~Abe {\it et al.}  [DOUBLE-CHOOZ Collaboration],
  Phys.\ Rev.\ Lett.\  {\bf 108}, 131801 (2012)

\bibitem{An:2012eh}
  F.~P.~An {\it et al.}  [DAYA-BAY Collaboration],
  Phys.\ Rev.\ Lett.\  {\bf 108}, 171803 (2012)
  [arXiv:1203.1669 [hep-ex]].

\bibitem{Ahn:2012nd}
  J.~K.~Ahn {\it et al.}  [RENO Collaboration],
  Phys.\ Rev.\ Lett.\  {\bf 108}, 191802 (2012)
  [arXiv:1204.0626 [hep-ex]].


\bibitem{smirnov}
A.~Y.~Smirnov, arXiv:hep-ph/0402264.

\bibitem{qlc}
H.~Minakata and A.Y.~Smirnov, Phys. Rev. D {\bf 70}, 073009 (2004).

\bibitem{raidal}
M.~Raidal, Phys.\ Rev.\ Lett.\  {\bf 93},  161801
(2004).




\bibitem{Zheng:2011uz}
  Y.-j.~Zheng and B.-Q.~Ma,
  Eur.\ Phys.\ J.\ Plus {\bf 127}, 7 (2012)
  [arXiv:1106.4040 [hep-ph]].


\bibitem{Zhang:2012xu}
  X.~Zhang and B.-Q.~Ma,
  Phys.\ Lett.\ B {\bf 710}, 630 (2012)
  [arXiv:1202.4258 [hep-ph]].





\bibitem{Fogli:2012ua}
  G.~L.~Fogli, E.~Lisi, A.~Marrone, D.~Montanino, A.~Palazzo and A.~M.~Rotunno,
  Phys.\ Rev.\ D {\bf 86}, 013012 (2012).

\bibitem{Li:2005ir}
  N.~Li and B.-Q.~Ma,
  Phys.\ Rev.\ D {\bf 71}, 097301 (2005)
  [hep-ph/0501226].

\bibitem{dihedral}
  A.~Blum, C.~Hagedorn and M.~Lindner,
  Phys.\ Rev.\ D {\bf 77}, 076004 (2008);
  A.~Blum, C.~Hagedorn and A.~Hohenegger,
  JHEP {\bf 0803}, 070 (2008);
  J.~E.~Kim and M.~-S.~Seo,
  JHEP {\bf 1102}, 097 (2011).


\bibitem{phenomenology}
See, e.g.,


  P.~H.~Frampton and R.~N.~Mohapatra,
  JHEP {\bf 0501},  025 (2005)
  [arXiv:hep-ph/0407139];
  N.~Li and B.-Q.~Ma,
  Phys.\ Rev.\  D {\bf 71}, 097301 (2005)
  [arXiv:hep-ph/0501226];
  S.~Antusch, S.~F.~King and R.~N.~Mohapatra,
  Phys.\ Lett.\  B {\bf 618},  150 (2005)
  [arXiv:hep-ph/0504007];
  H.~Minakata,
  arXiv:hep-ph/0505262;
  J.~Ferrandis and S.~Pakvasa, Phys. Rev. D {\bf 71}, 033004 (2005);
S.K.~Kang, C.S.~Kim and J.~Lee, Phys. Lett. B {\bf 619}, 129 (2005);
S.~Antusch, S.F.~King and R.N.~Mohapatra, Phys. Lett. B {\bf 618},
150 (2005); M.A.~Schmidt and A.Y.~Smirnov, Phys. Rev. D {\bf 74},
113003 (2006);
  J.~E.~Kim and J.~-C.~Park,
  JHEP {\bf 0605}, 017 (2006);
  K.~A.~Hochmuth and W.~Rodejohann,
  Phys.\ Rev.\  D {\bf 75}, 073001 (2007)
  [arXiv:hep-ph/0607103];
F.~Plentinger, G.~Seidl and W.~Winter, Phys. Rev. D {\bf 76}, 113003
(2007);

  G.~Altarelli, F.~Feruglio and L.~Merlo,
  JHEP {\bf 0905}, 020 (2009);
  C.~H.~Albright, A.~Dueck and W.~Rodejohann,
  Eur.\ Phys.\ J.\ C {\bf 70}, 1099 (2010).

\bibitem{Zheng10}
Y.-j.~Zheng, Phys.\ Rev.\ D {\bf 81}, 073009 (2010) [arXiv:1002.0919
[hep-ph]].

\bibitem{Zhang:2012zh}
  X.~Zhang, Y.-j.~Zheng and B.-Q.~Ma,
  Phys.\ Rev.\ D {\bf 85}, 097301 (2012) [arXiv:1203.1563 [hep-ph]].


\bibitem{triminimal}
S.~Pakvasa, W.~Rodejohann and T.J.~Weiler, Phy. Rev. Lett. {\bf
100}, 111801 (2008).





\bibitem{He:2008td}
  X.-G.~He, S.-W.~Li and B.-Q.~Ma,
  Phys.\ Rev.\ D {\bf 78}, 111301 (2008)
  [arXiv:0809.1223 [hep-ph]].

\bibitem{He:2009jm}
  X.-G.~He, S.-W.~Li and B.-Q.~Ma,
  Phys.\ Rev.\ D {\bf 79}, 073001 (2009)
  [arXiv:0903.2880 [hep-ph]].

\bibitem{Li:2008aa}
  S.-W.~Li and B.-Q.~Ma,
  Phys.\ Rev.\ D {\bf 77}, 093005 (2008)
  [arXiv:0806.0670 [hep-ph]].

\bibitem{Qin:2011ub}
  N.~Qin and B.-Q.~Ma,
  Phys.\ Rev.\ D {\bf 83}, 033006 (2011)
  [arXiv:1101.4729 [hep-ph]].


\bibitem{Li:2005yj}
  N.~Li and B.-Q.~Ma,
  Eur.\ Phys.\ J.\ C {\bf 42}, 17 (2005)
  [hep-ph/0504161].



\bibitem{Qin:2011bq}
  N.~Qin and B.-Q.~Ma,
  Phys.\ Lett.\ B {\bf 702}, 143 (2011)
  [arXiv:1106.3284 [hep-ph]].


\bibitem{Ahn:2011ep}
  Y.~H.~Ahn, H.-Y.~Cheng and S.~Oh,
  Phys.\ Lett.\ B {\bf 715}, 203 (2012) arXiv:1105.4460 [hep-ph];  
  Phys.\ Rev.\ D {\bf 84}, 113007 (2011)  [arXiv:1107.4549 [hep-ph]].  

\bibitem{Ma:2012zm}
  B.-Q.~Ma,
  arXiv:1205.0766 [hep-ph].

\bibitem{Chauhan:2006im}
  B.~C.~Chauhan, M.~Picariello, J.~Pulido and E.~Torrente-Lujan,
  Eur.\ Phys.\ J.\ C {\bf 50}, 573 (2007)
  [hep-ph/0605032].

\bibitem{Datta:2005ci}
  A.~Datta, L.~Everett and P.~Ramond,
  Phys.\ Lett.\ B {\bf 620}, 42 (2005)
  [hep-ph/0503222].

\bibitem{King:2012vj}
  S.~F.~King,
  arXiv:1205.0506 [hep-ph].

\bibitem{Antusch:2012fb}
  S.~Antusch, C.~Gross, V.~Maurer and C.~Sluka,
  arXiv:1205.1051 [hep-ph].


\bibitem{Masina}
  I.~Masina,
  Phys.\ Lett.\ B {\bf 633}, 134 (2006)  [hep-ph/0508031].  





\bibitem{Everett:2005ku}
  L.~L.~Everett,
  Phys.\ Rev.\ D {\bf 73}, 013011 (2006)
  [hep-ph/0510256].


\bibitem{Li:2004nh}
  N.~Li and B.~-Q.~Ma,
  Phys.\ Lett.\ B {\bf 600}, 248 (2004)
  [hep-ph/0408235].

\bibitem{Rodejohann:2003sc}
  W.~Rodejohann,
  Phys.\ Rev.\ D {\bf 69}, 033005 (2004).


\bibitem{Li:2004dn}
  N.~Li and B.-Q.~Ma,
  Phys.\ Rev.\ D {\bf 71}, 017302 (2005)
  [hep-ph/0412126].


\bibitem{King:2007pr}
  S.F.~King,
  Phys.\ Lett.\ B {\bf 659}, 244 (2008).

\end{thebibliography}
\end{document}